\documentstyle[aps,twocolumn]{revtex}
\author{Jeferson J. Arenzon, Mario Nicodemi$^*$ and Mauro Sellitto$^*$}
\address{
Dipartimento di Scienze Fisiche \\ Universit\`a di Napoli
`Federico II'
\\ Pad. 19 -- Mostra d'Oltremare \\ 80125 -- Napoli -- ITALY
\vspace{0.5cm}
\\ $*$ Unit\`a INFM e Sezione INFN di Napoli 
\vspace{0.5cm} 
}
\title{Equilibrium properties of the Ising frustrated lattice gas}
\date{{\bf J. Physique I}, 1996 to be published}

\newcommand{\tr}{\mathop{\rm Tr}}
\newcommand{\Dz}{{\cal D}z\:}

\newcommand{\ponto}{\;\;.}
\newcommand{\e}{\mbox{e}}
\newcommand{\Dzo}{{\cal D}z_0\:}
\newcommand{\Dzu}{{\cal D}z_1\:}
\newcommand{\bea}{\begin{eqnarray*}}
\newcommand{\eea}{\end{eqnarray*}}

\begin{document}
\maketitle

\begin{abstract}
We study the equilibrium properties of an Ising frustrated lattice 
gas with a mean field replica approach. 
This model bridges usual {\em Spin Glasses} and a version of {\em 
Frustrated 
Percolation} model, and has proven relevant to describe the glass 
transition. It shows a rich phase diagram which in a definite 
limit reduces to the known Sherrington-Kirkpatrick spin glass
model. 
\end{abstract}

\section{Introduction}

The Ising Spin Glass transition has been discovered to describe many 
seemingly different phenomena, and to model real systems much 
beyond what originally thought \cite{MezardParisiVirasoro}. 
The idea of introducing frustrated Hamiltonians 
to capture the essential physics of glasses, 
random media properties, evolutionary models,
protein and {\small RNA} folding, granular packing, 
dynamics of complex flow, 
and many others, has grown fertile 
in the last years 
\cite{Kirk,Kur,LAS,Edwards,Coniglio,Nicodemi,ConiglioHerrmann,Peliti,Higgs}.

Recently, in this panorama, an Ising Spin Glass like model, a general version 
of the frustrated lattice gas, has been introduced for its new interesting 
Monte Carlo dynamical and equilibrium features relevant to the description 
of the glass transition \cite{Coniglio,Nicodemi}, and with some 
variations has been applied in the context of phase transitions in granular 
packing \cite{ConiglioHerrmann}. 

In this paper we study the mean field equilibrium properties 
of such a model, adopting standard replica formalism. 
The system we consider is characterized by the following Hamiltonian:
\begin{equation}
{\cal H} = J \sum_{\langle ij\rangle} (1-\varepsilon_{ij}S_iS_j)n_in_j
-\mu \sum_i n_i -h\sum_i S_in_i 
\label{H}
\end{equation}
where the lattice gas site variables $n_i=0,1$ have an Ising internal 
degree of freedom $S_i=\pm 1$ and $\langle ij\rangle$ denotes summation
over all nearest-neighbor pairs of sites. The $\varepsilon_{ij}=\pm 1$ are 
quenched random 
variables, $h$ is a magnetic field  applied on the system 
and $\mu$ is a chemical potential  for the site variables.
Essentially, the model considers a
lattice gas in a frustrated medium where the particles have an internal 
degree of freedom (given by its spin) that accounts, for example, for  
possible orientations of complex molecules in glass forming liquids.
As stressed by Coniglio \cite{Coniglio}, these 
steric effects are greatly responsible for the 
geometric frustration appearing in glass forming
systems at low temperatures or high densities.
As is detailed in ref.\cite{Nicodemi}, this model offers a
clear and intuitive picture of the mechanism leading to a glass transition,
qualitatively reproducing the complex dynamical behaviour present
in this regime. 

The presented Hamiltonian is a natural bridge 
between {\em Frustrated Percolation} \cite{Coniglio}
and standard {\em Ising Spin Glasses} (SG). 
Indeed, this two apparently different models are obtained  
in two definite limits of its parameters. If $\mu\rightarrow\infty$ 
and $J/\mu\rightarrow 0$, for energetical reasons each site must be 
filled and the known Ising Spin Glass is obtained; on the contrary 
if $J\rightarrow\infty$ and $\mu/J\rightarrow 0$, generally even at $T=0$ 
the configuration with each site filled is impossible. In this last limit 
only site configurations which do not close ``frustrated loops", 
i.e. loops of filled sites whose spins are not satisfying all 
mutual interactions, are allowed. 
This corresponds to a site Frustrated Percolation in which clusters have a 
further weight factor of $2^{N_c}$ ($N_c$ is the number of clusters in the 
system).
Moreover, introducing clusters \`a la Kasteleyn and Fortuin, this Hamiltonian 
may be described in terms of a site-bond correlated and frustrated 
percolation.

With a simple transformation, $\tau_i\equiv S_in_i$, this model may 
be changed into 
an Ising spin-1 Blume-Emery-Griffiths model (BEG) \cite{BEG} in which only 
the
bilinear coupling is affected by the quenched disorder $\varepsilon_{ij}$, 
while the spin biquadratic term has a coupling with opposite sign in
relation to the original BEG model \cite{HostonBerker} 
(in notation of \cite{HostonBerker}, $K/J=-1$) and no 
disorder. In the case without frustration ($\varepsilon_{ij}=1$), that
has also been studied in \cite{HostonBerker},
the order parameters are the 
diluted magnetization $m=\langle Sn \rangle$ and the particles density 
$d=\langle n\rangle$.
In the $\mu\to\infty$ limit one recovers the Ising model, with $d=1$.
In mean field, at $T=0$, $d=m=\Theta(\mu)$ if $\mu\neq0$ and $d=m=1/3$
for $\mu=0$. This point with density $0<d<1$ will become an
interval when frustration is introduced. 
Moreover, expanding for small $m$ we obtain the equation satisfied by
the critical temperature
\begin{equation}
\frac{J}{T_c} = 1+\exp\left(1-\frac{\mu}{T_c}\right) 
\label{TcFer}
\end{equation}
and one can see that this transition line is reentrant 
\cite{HostonBerker},
effect that will also disappear when introducing frustration.

In the following sections we study the replica mean field theory of
Hamiltonian (\ref{H}) with a gaussian distributed coupling. 
The phase diagram of the model presents several interesting regions 
depending on the values of $T$ and
$\mu/J$. For highly negative values of $\mu/J$, there
is only a paramagnetic phase. Lowering the temperature at small negative 
values of $\mu/J$, 
the system has a discontinuous transition to a spin glass phase,
in which even at zero temperature the density is lower 
than one. Increasing $\mu/J$, the spin glass 
transition becomes continuous, while the zero temperature density
is still below one. This occurs up to a certain point where 
the density becomes unity (at $T=0$) signalling that we are approaching 
the Sherrington-Kirkpatrick limit. The Parisi replica symmetry breaking 
solution seems to hold whenever a spin glass transition is encountered.

\section{Mean Field Results}

We present in this section results obtained in the mean field approximation. 
The starting point is the calculation of the free energy $f$ according 
with the replica trick \cite{MezardParisiVirasoro}:
\begin{equation}
 \beta f = -\lim_{n \rightarrow 0} \frac{\ln [ Z^n ]_{av.} }{nN}
\end{equation}
where $[ \ldots ]_{av.}$ stands for the 
average over the disorder, which we suppose gaussian with zero mean
and variance $J^2/N$. We obtain:
\begin{eqnarray}
\beta f &=& \lim_{n\to 0} \frac{1}{n} \left\{ \frac{\beta^2 J^2}{2}
\sum_{a<b} q_{ab}^2 + 
 \frac{\beta J}{2}\left( \frac{\beta J}{2}-1\right)
\sum_a d_a^2 \right. \nonumber \\ && \left. 
-\ln \tr_{\{n^a,S^a \}} \e^{-\beta {\cal H}_{eff}} \right\}
\end{eqnarray}
where the single site Hamiltonian is:
\begin{eqnarray}
{\cal H}_{eff} &=& -\beta J^2 \sum_{a<b} q_{ab} S^a n^a S^b n^b
- J\left( \frac{\beta J}{2}-1\right) \sum_a d_an^a \nonumber \\
 && - \mu\sum_a n^a -  h \sum_a S^an^a
\end{eqnarray}

The self consistent equations for the order parameters are given
by the saddle points of $f$ and read:
\begin{eqnarray}
q_{ab}&=&\langle S^an^aS^bn^b \rangle   \\
d_a&=&\langle n^a \rangle
\end{eqnarray}
where the average is done using the effective Hamiltonian.
The overlap $q_{ab}$ has a certain
degree of dilution in respect to the parameter introduced in the SK model
\cite{SK} and reduces to it in the limit $d_a=1$.

\subsection{Replica Symmetry}

To get a general qualitative picture of the phase diagram of the system, 
we first made a simple replica symmetric (RS) assumption, that is, 
$q_{ab} = q(1-\delta_{ab})$ 
and $d_a = d$. The free energy then reads:
\begin{eqnarray}
\beta f_s = &-&\frac{1}{4}\beta^2 J^2 (q^2-d^2) -\frac{1}{2}
\beta J d^2 -\ln 2  \\ &-& 
\int\Dz \ln\left[ 1 + \e^{\Xi}\cosh(\beta J \sqrt{q}z
+\beta h)\right] \nonumber
\end{eqnarray}
where the gaussian measure is
${\cal D}z \equiv \frac{dz}{\sqrt{2\pi}} \e^{-z^2/2} \nonumber$ and
\begin{equation}
\Xi \equiv \frac{\beta^2 J^2}{2} (d- q) +\beta (\mu-Jd)
\label{theta}
\end{equation}
The saddle point equations obeyed by the order parameters are:
\begin{equation}
d=\int\Dz\frac{\cosh(\beta J\sqrt{q}z+\beta h)}{\e^{-\Xi} +
\cosh(\beta J\sqrt{q}z+\beta h)} 
\label{d}
\end{equation}
and
\begin{equation}
q = \int\Dz   \frac{\sinh^2(\beta J\sqrt{q}z+\beta h)}{
[\e^{-\Xi} + \cosh(\beta J\sqrt{q}z+\beta h)]^2} 
\label{q}
\end{equation}
The effects introduced by the magnetic field will in general not be 
considered here and in what follows
we take, unless mentioned, $h=0$.

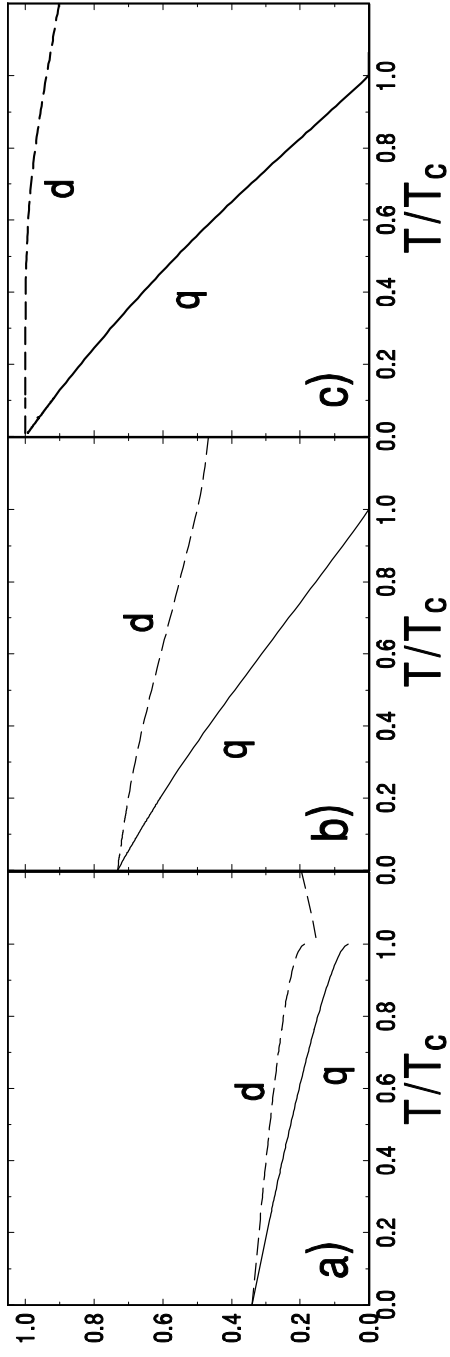
\begin{figure}[h]
\caption{The order parameters $q$ and $d$ for several values of $\mu=
-0.65$ (a), 0 (b) and 3 (c). 
Notice the diverse behaviour of $d$ (and $q$) as $T\to 0$ and the
discontinuous transition in (a). The respective critical temperatures
are $T_c\simeq 0.13$ (a), 0.5 (b) and 0.94 (c).} 
\label{qdmu}
\end{figure}

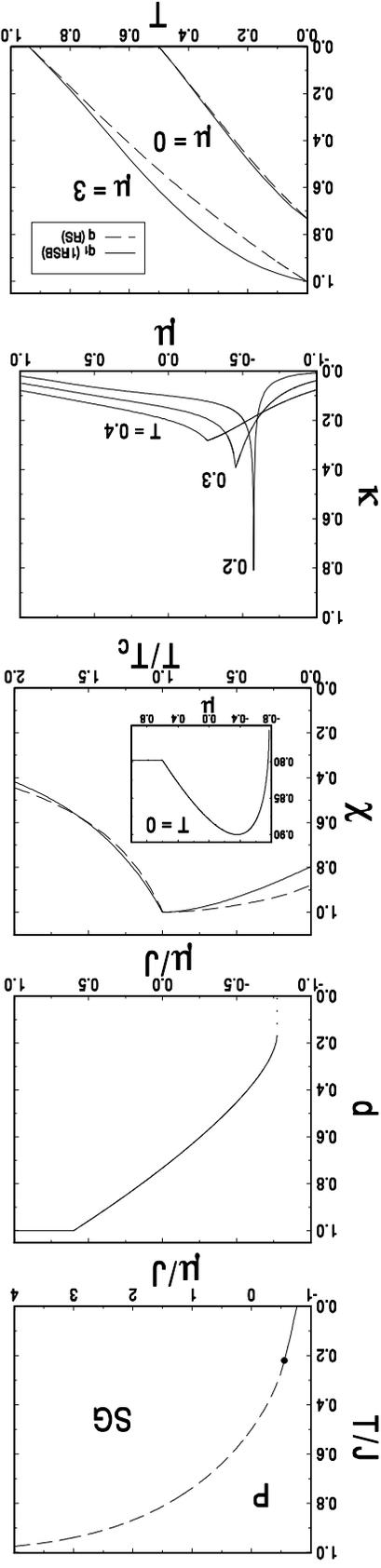
\begin{figure}[h]
\caption{The phase diagram $T$ versus $\mu$. The dashed line stands for
the continuous transition while the solid line corresponds to the first 
order one ($-0.77<\mu<-0.56$) and both meet at a tricritical point.
In particular, when $J=-2\mu$, $T_c=J(1+e)^{-1}\simeq 0.27 J$
and when $\mu=0$, $T_c=J/2$.} 
\label{phasediagram}
\end{figure}

To characterize the system, in 
fig.\ref{qdmu} we present some representative curves for both
$q$ and $d$ for several values of $\mu/J$. For large values,
 the system approaches the SK limit ($T_c\to J$, $d\to1$ and 
$q\to q_{SK}$). For $\mu/J>-0.56$,
the system has a continuous transition ($q\sim 0$)  at $T_c$
satisfying
\begin{equation}
\frac{J}{T_c} = 1+\exp\left(1-\frac{J}{2T_c} -\frac{\mu}{T_c}\right)
\end{equation}
value to be compared with eq.(\ref{TcFer}).
Decreasing $\mu/J$ further, the transition line becomes first order
(for $-0.77<\mu/J<-0.56$) as signaled by a jump in the order parameter.
At the transition line a partial freezing takes place ($q<d<1$), 
behaviour that has to be compared with other disordered models with 
discontinuous transitions like, for instance, the $p$-states Potts glass
with $p>4$ \cite{Potts} and the $p$-spin interaction model with $p>2$
\cite{p-spin}.
We can also
see from fig.\ref{qdmu} that at low temperature $q$ approaches $d$,
while the actual value they assume at $T=0$, where $q=d$ (the system being 
fully frozen), depends 
on the chemical potential as can be seen in fig.\ref{dt0}.
 The point where the transition changes behavior
($\mu/J\simeq -0.56$) turns 
out to be a tricritical one when including a non zero magnetic field.
The system is a simple paramagnet ($q=0$) below this region.
 This information is summarized in the phase diagram $T$ versus 
$\mu$ presented in fig.\ref{phasediagram}.
The reentrant phase found in the case without frustration \cite{HostonBerker}
is replaced here by these various regions.

It is interesting to study the $T=0$ limit of this model.
For $\mu$ above the point
\begin{equation}
\frac{\mu^*}{J} = 1 -\frac{1}{\sqrt{2\pi}} \simeq 0.6
\label{mustar}
\end{equation}
it is possible to see that $d=1$ and $C\equiv\beta J (d-q)=\sqrt{2/\pi}$, 
results known to be characteristic of the Sherrington-Kirkpatrick SG. 
Something changes below $\mu^*$, eq.(\ref{mustar}), where 
the saddle point equations give:
\begin{eqnarray}
d&=&\mbox{erfc} \left( \frac{-\frac{1}{2} C+d-\frac{\mu}{J}}{\sqrt{2d}} \right)
\\
C&=&
\sqrt{\frac{2}{d\pi}} \exp\left[ -\frac{\left(-\frac{1}{2}C+d-
\frac{\mu}{J}\right)^2}{2d}\right] 
\end{eqnarray}
These equations imply $d<1$ at $T=0$, as it should be in the true 
$J\rightarrow\infty$ limit. 
On the other hand, below $\mu_c\simeq -0.77$,
just the paramagnetic solution $q=d=0$ is present,
the transition being discontinuous (see fig.\ref{dt0}), in accordance
with the above phase diagram.
The main novelties present in the model appear in this region where the
chemical potential is sufficiently low and  frustrated
loops are not completely occupied ($d<1$).  
The appearance of this region is an effect introduced by the disorder 
since in the no frustrated
case, it reduces to the point $\mu=0$.

The entropy per site is
\begin{equation}
s = \frac{1}{2} \beta^2 J^2 (q^2-d^2) + \frac{1}{2}\beta Jd^2-\beta\mu d 
-\beta f_s \ponto
\end{equation}
As $T\to\infty$, $s\to 2\ln 2$ since our phase space has $4^N$ possible
states, and at $T=0$:
\begin{equation}
s_0=-\frac{1}{4} C^2 +(1-d)\ln 2 \ponto
\label{entropyt0}
\end{equation}
For $\mu>\mu^*$, $s_0=-1/2\pi$, while for $d=0$, $s_0=\ln 2$.
Here we clearly see the signal of instability of the replica symmetric
solution. The next section
treats the first step of replica symmetric breaking in the
Parisi scheme, leading to corrections to these results.
We also obtain the point where $s_0=0$, that is, $\mu_0\simeq -0.026$.
The reason which makes positive the replica symmetric 
entropy  below this value is the presence of free spins when
the density is lower than unity. In
those sites where $n_i=0$, the spins are free to assume any
orientation, and this increases the entropy (second term in 
eq.(\ref{entropyt0})).
The free energy hessian eigenvalues have to be calculated in order to verify 
the stability of the replica symmetry solution.
 Doing so  we obtain from the condition of
positive eigenvalues, the AT line \cite{AT}
\begin{equation}
\frac{1}{\beta^2 J^2} > \int \Dz \frac{\left[
1+\e^{-\Xi}\cosh(\beta J\sqrt{q}z)\right]^2}{\left[
\e^{-\Xi}+\cosh(\beta J\sqrt{q}z)\right]^4}
\end{equation}
below which the replica symmetric solution is unstable. It can be  verified
that the above equation is satisfied nowhere below $T_c$ and, as 
in the SK model, here too the replica symmetric solution is unstable,
although the ``degree of stability" may vary with $\mu$.

The susceptibility, $\chi = \beta (d-q)$, presents a cusp at $T_c$,
as can be seen in fig.\ref{susc}. The zero temperature value of
$\chi$, $\chi_0=C/J$, depends on $\mu$ as shown in the inset of 
fig.\ref{susc}. Above $\mu^*$ it has a constant value, while below
it presents a maximum. 
As expected above $T_c$, $\chi=d/T$, depending just on the density.
We might have chosen to apply $h$ to all spins in eq.(\ref{H})
whether their sites were occupied or not ($h\sum_iS_i$). In this case 
in the region where $d(T=0)<1$, the low temperature
susceptibility (now $\chi = \beta (1-q)$) would diverge as 
$T^{-1}$ due to the free spins which have a strong response even 
to a weak field.

We report here the values of other quantities in order
to characterize the system and 
for comparison with the known SG. The internal energy per spin is
\begin{equation}
u=\frac{1}{2} \beta J^2(q^2-d^2) +\frac{J}{2} d^2 -d\mu 
\end{equation}
and, at $T=0$,
$$
u_0 = -JCd+\frac{J}{2} d^2-d\mu \ponto
$$
The compressibility $\kappa=\beta^{-1}\partial d /\partial \mu$ 
has a cusp when the transition is continuous, although
presents a divergence when the transition is first order, 
as can be seen in fig.\ref{compre}.

\subsection{Replica Symmetry Breaking}

Results in the previous section have shown the instability of the RS solution 
and we report here the first Parisi correction to it. However, it will appear
 that the phase diagram sketched above in RS is 
not altered. 

Following the Parisi scheme \cite{MezardParisiVirasoro}, in the first 
step of replica symmetry breaking (1RSB) the $n$ replicas are
divided in $n/m$ blocks containing $m$ replicas. 
Different replicas in the same block have overlap $q_1$ while
those in different blocks have overlap $q_0$.
 Thus, the 1RSB free energy is:
\begin{eqnarray}
\beta f_1 = &-&\frac{1}{4} \beta^2 J^2 \left[ (1-m) q_1^2 + m q_0^2 -d^2
\right] - \frac{1}{2} \beta J  d^2 
 \nonumber \\ 
&-& \ln 2 -\frac{1}{m} \int \Dzo \ln \int \Dzu A^m(z_0,z_1)
\end{eqnarray}
where:
\bea 
A(z_0,z_1) &\equiv& 1 +  
\e^{\Xi_1} \cosh (\beta h + \beta J z_0 \sqrt{q_0} + \beta J z_1 
\sqrt{q_1-q_0})  \\
\Xi_1     &\equiv& \frac{\beta^2 J^2}{2} (d-q_1) +\beta (\mu-Jd) 
\eea

As usual in spin-glass theory, we have to maximize the free energy as 
a function of $q$, $d$ and $m$, and the saddle point equations are:
\begin{eqnarray}
d &=& 1- \int\Dzo [A^{-1}]_A \label{d1} \\
q_0 &=& \int\Dzo [B]_A^2 \label{q0} \\
q_1 &=& \int\Dzo [B^2]_A \label{q1}
\end{eqnarray}
($d\geq q_1\geq q_0$) and $m$ satisfies
\begin{eqnarray}
&& \frac{1}{4} m^2 \beta^2 J^2 (q^2_1 - q^2_0) 
 -m\int\Dzo [\ln A]_A  \nonumber \\
&& + \int\Dzo\ln \int \Dzu A^m = 0
\label{m}
\end{eqnarray}
Here we have defined:
\begin{equation}
B(z_0,z_1) \equiv \frac{\sinh (\beta h + \beta J z_0 \sqrt{q_0} 
+ \beta J z_1 \sqrt{q_1-q_0})}{  \e^{-\Xi_1} A(z_0,z_1)}
\end{equation}
and
\begin{equation}
[X]_A = \frac{\int \Dzu A^m X }{ \int \Dzu A^m} \ponto
\end{equation}

The numerical solutions, which can be obtained by either maximizing the free
energy or by solving the coupled system of saddle point equations,
are shown in fig. \ref{q0q1d}.
The density does not change in relation to the replica symmetric
case, while the effect of replica symmetry breaking
on $q_1$ is rather small, like in the SK model.
For $m$, the behavior is analogous to the one found by Parisi in the
SK model, $m=0$ at both $T=0$ and $T=T_c$, and have a maximum at an
intermediate temperature.

In analogy with the SK model, the difference between $f_s$ and $f_1$
is only perceptible at low temperatures, although depending on the
value of $\mu$, this effect becomes more difficult to notice. It
can also be seen that $f_1\ge f_s$, as expected, and that the
regions with wrong curvature (negative entropy) in the RS solution
are largely reduced.

The susceptibility is given by
\begin{equation}
\chi = \beta [d+(m-1)q_1 - mq_0]
\end{equation}
and the results of 1RSB are higher than the ones found in RS (see
fig.\ref{susc}), as in
the SK model. 
We have also checked that, at $T=0$, the point where the density becomes 
lower than one coincides with the one obtained with replica symmetry, 
eq.(\ref{mustar}). 

Our calculations in 1RSB confirm that the structure of the 
phase diagram found in replica symmetry is not altered when breaking it.
The first order transition is quite similar to the one found in 
other spin glass models \cite{REM,Potts,p-spin,spherical}, 
although the model deserves a more detailed study to clarify 
whether the transition is or not exactly described by one step
of replica symmetry breaking. In this case the glassy behavior of
the model could be rather peculiar and different from other disordered
system with discontinuous transition.

\section{Summary and Conclusions}

In brief, we have studied the mean field theory of a simple spin glass 
version of the 
Blume-Emery-Griffiths model with standard replica formalism.
This model, described by the Hamiltonian (\ref{H}), bridges 
the usual SG, which is obtained in the 
limit ($J/\mu\rightarrow 0$, $\mu\rightarrow\infty$), and 
a version of ``frustrated percolation", which is recovered if 
($\mu/J\rightarrow 0$, $J\rightarrow\infty$).

We have seen that the introduction of site variables $n_i$
enrich the phase diagram of the spin glass. 
Specifically we have shown that the SG phase around 
the  ``frustrated percolation'' limit, i.e. $\mu/J=0$, has 
peculiarities signaled, for instance, by a site density below one even 
in the $T=0$ limit. 
Moreover, a discontinuous transition appears for 
negative enough values of $\mu/J$, which may be relevant to the 
description of the structural glass
transition \cite{Kirk}.
 These two effects, namely
the broad $d<1$ region around $\mu=0$ and the discontinuous transition,
are not present in the unfrustrated model 
where a reentrant phase was present \cite{HostonBerker}.

The transition is characterized by a cusp in the susceptibility,
although for the compressibility there is a cusp when the
transition is continuous and a divergence when it is first
order.
By comparing the results obtained with 1RSB with those found assuming
symmetry, we see that we are already very near the true mean
field behavior of the system, since the corrections introduced are
small. 
On the other hand, a more detailed analysis should be done
in the region where the transition is discontinuous in order
to verify whether, as happens in other models with
discontinuous transition \cite{REM,Potts,p-spin,spherical}, 
the 1RSB solution is exact. 

The mean field results
obtained here are in qualitative
accordance with those found through numerical simulation 
for $D=3$ \cite{Nicodemi}, although a
 simulation for the infinite range version is still lacking
and it would be welcomed.

It would be interesting to study the effect 
of varying the couplings parameters
in Hamiltonian eq.(\ref{H})\cite{nuovo}, 
as done in the non frustrated case in \cite{BEG,HostonBerker} where 
qualitatively different phase diagrams are so found. Also, by introducing
a non zero mean in the gaussian distribution, it might be possible
to study the appearance of the reentrant phase and tune the
length of the zero temperature $0<d<1$ interval \cite{nuovo}. 
We expect, by the above results, that the interplay of connectivity
and frustration would lead to a richer static (and dynamic)
behavior than the usual spin glasses.

\vskip 1\baselineskip
\noindent
{\large\bf Acknowledgments}: We thank A. Coniglio and L.
Peliti for interesting discussions and for a 
careful reading of the manuscript. 
JJA acknowledges partial support from the brazilian agency CNPq.

\begin{figure}[h]
\caption{The density $d$ versus $\mu$ at $T=0$. The point for which 
$d\to 1$ is $\mu=1-(2\pi )^{-1/2}$ while for $\mu<\mu_c\simeq -0.77$ 
there is only the $d=0$ solution.} 
\label{dt0}
\end{figure}

\begin{figure}[h]
\caption{The magnetic susceptibility $\chi$ versus $T$ for
$\mu=0$ and 3 in RS. Inset: $T=0$ susceptibility as a
function of $\mu$, showing a constant value above $\mu^*$
and a maximum in the region where $d(T=0)<1$.}
\label{susc}
\end{figure}

\begin{figure}[h]
\caption{The compressibility $\kappa$ as a function of $\mu/J$
for $T=0.4$, 0.3 and 0.2. In the first two cases the transition is
continuous and $\kappa$ presents a cusp while in the last one 
it has a divergence since the transition is discontinuous.} 
\label{compre}
\end{figure}

\begin{figure}[h]
\caption{The order parameter $q_1$  obtained with
1RSB and $q$ with RS for $\mu=0$ and 3.
Regarding $d$, we did not find any difference between the RS
and 1RSB solutions, while for $q_1$, the correction
is small.} 
\label{q0q1d}
\end{figure}

\end{document}